\documentclass[aps, pra, 10pt, twocolumn, superscriptaddress,floatfix,longbibliography]{revtex4-1}
\usepackage{bbold}
\usepackage{graphicx}
\usepackage{amsmath,amssymb,amsfonts}
\usepackage[percent]{overpic}
\usepackage{braket}
\usepackage{bm}
\usepackage[breaklinks=true,colorlinks,citecolor=blue,linkcolor=blue,urlcolor=blue]{hyperref}
\usepackage[usenames, dvipsnames]{color}
\usepackage{mathrsfs}
\usepackage{comment}
\usepackage{orcidlink}

\newcommand{\dstate}{\rho} 

\newcommand{\ham}{H}

\newcommand{\ergo}{\mathcal{E}}
\newcommand{\destr}{b}
\newcommand{\constr}{b^{\dagger}}
\newcommand{\exch}{\sigma_{eg}}
\newcommand{\dexch}{\sigma_{ge}}

\newcommand{\Tr}{\text{Tr}}

\begin{document}
\title{Many-body enhancement of energy storage in a waveguide-QED quantum battery}
\author{Salvatore Tirone\orcidlink{0000-0002-4880-4329}}
\email{s.tirone@uva.nl}
\affiliation{Scuola Normale Superiore, I-56126 Pisa, Italy}
\affiliation{QuSoft, Science Park 123, 1098 XG Amsterdam, the Netherlands}
\affiliation{Korteweg--de Vries Institute for Mathematics, University of Amsterdam, Science Park 105-107, 1098 XG Amsterdam, the Netherlands}
\author{Gian Marcello Andolina\orcidlink{0000-0002-4219-7177}}
\affiliation{JEIP, UAR 3573 CNRS, Collège de France, PSL Research University, 11 Place Marcelin Berthelot,  F-75321 Paris, France}
\author{\\ Giuseppe Calaj\`o\orcidlink{0000-0002-5749-2224}}
\affiliation{Istituto Nazionale di Fisica Nucleare (INFN), Sezione di Padova, I-35131 Padova, Italy}
\author{Vittorio Giovannetti}
\affiliation{Scuola Normale Superiore, I-56126 Pisa, Italy}
\author{Davide Rossini\orcidlink{0000-0002-9222-1913}}
\affiliation{Dipartimento di Fisica dell’Università di Pisa and INFN, 
Largo Pontecorvo 3, I-56127 Pisa, Italy}

\begin{abstract}
Quantum batteries have demonstrated remarkable charging properties, showing that a quantum advantage is possible in the realm of quantum thermodynamics. However, finding an effective strategy to store energy for long periods remains crucial in these systems.
Here, we investigate different configurations of a waveguide-QED system acting as a quantum battery and show that, in this context, collective effects can slow down the self-discharging time of the battery, thus improving the storage time. Specifically, when the artificial atoms of the array are arranged randomly, the energy and ergotropy of the optical system are shown to decay at a subexponential rate over long periods, in contrast to the energy decay of a single atom in a waveguide. In the case of atoms arranged in an ordered lattice, collective effects slow down dissipative discharging only for a specific lattice spacing. Thus, in both configurations, collective effects can be used to boost the energy-protection properties of optical systems.
\end{abstract}


\maketitle

\section{Introduction}

In the last decade, quantum technologies have been a topic of intense research and development. These technologies take advantage of the unique properties of quantum physics, such as coherence and entanglement, for practical applications. This has led to significant advances in areas like quantum computing~\cite{Preskill_2018}, sensing~\cite{Degen_2017}, communication, and simulation~\cite{Georgescu_2014}. However, until recently~\cite{EnergyInitiative}, the crucial area of energy production and management has been only partially addressed within the realm of quantum technologies.  As an example, quantum memories have been extensively investigated within the field of quantum information as devices capable of storing quantum states, while preserving their fragile quantum properties \cite{Fleischhauer2000, Fleischhauer2002, QMrev2009}. 
In contrast, the concept of quantum batteries (QBs)—devices where energy can be injected, stored, and eventually recovered~\cite{qbcolloquium}—has only recently garnered increasing interest within the field of quantum thermodynamics.

A QB~\cite{Campaioli2018,qbcolloquium} is composed of multiple identical units, also called quantum cells, often modeled as two-level systems (TLSs). Energy can be stored in these units by exciting them to a higher energy state. Such energy can then be released by allowing the units to relax back to their lower energy state, with the energy difference being used to perform work.
The interest in these systems stems from the fact that entanglement (or many-body interactions) between the units of the battery could allow for a super-extensive charging power~\cite{Alicki13,Hovhannisyan13,Binder15,Campaioli17,JuliFarr2020,Andolina2019,Farina2019,rossini2019quantum,Rossini2019,Gyhm2022} where the battery could be charged faster collectively. Due to this property, most previous studies mainly focused on the charging step, with the aim of increasing the charging power by means of collective effects. 

However, a truly efficient QB should be capable of more than just quick charging; it must also be able to store energy for a significant amount of time. In fact, a quantum device that can absorb energy quickly but fails to retain it for a long period does not fit the definition of a ``battery", but rather, it might be more appropriate to term it as an ``absorber".  
Various works in the literature deal with the storage phase~\cite{qwc, cap_nonloc, role_coherence, Bai_2020, Tabesh_2020, Ghosh_2021, Santos_2021, Zakavati_2021, Landi_2021, Morrone_2022, Sen_2023,Liu2019, PhysRevE.100.032107, PhysRevApplied.14.024092, PhysRevE.101.062114, PhysRevResearch.2.013095, Goold2016, liu2021boosting, PhysRevE.105.054115, Fischer2024, lu2024topologicalquantumbatteries, Downing_2024, PRXQuantum.5.030319, Bai2024}. Our approach, however, introduces a novel direction by utilizing the collective effects of an optical system that exhibits many-body localization properties alongside other distinctive features, as elaborated in the following. Moreover, this system not only holds the potential for developing a QB with a large number of sites, but it is also well-suited for experimental investigation in a waveguide-quantum electrodynamics (QED) setup~\cite{roy_rev_mod, Darrick_rev_mod}, offering a realistic pathway to test and validate our results.

In this paper, we consider the experimentally relevant platform of waveguide QED, which has emerged in the latest years as a powerful and promising light-matter interface.
Waveguide QED broadly refers to scenarios where one or more quantum emitters are coupled to the light confined into a one-dimensional photonic channel~\cite{shen_fan_prl, roy_rev_mod, Darrick_rev_mod, rev_wqedarray}.  This setting has been implemented in the optical regime, with cold atoms or quantum dots coupled to optical fibers and photonic crystal waveguides~\cite{review_lodahl,tiranov2023collective,Hood2016,Prasadnat,Laurat}, and at microwave frequencies with superconducting circuits, where artificial atoms 
are strongly coupled to on-chip microwave resonators or transmission lines~\cite{Astafiev2010,painter2,Kannan,ustinov}.
The 1D confinement allows for highly controlled interactions between light and matter, which can be used for various quantum technological applications, such as the dissipative preparation of atomic entangled states~\cite{Stannigel2012,Pichler2015}, the preparation of many-body states of light~\cite{Harold_bs,mahmo_calajo,calajo2022emergence,pichler2017universal,Painter_clusterstate,mahmoodian2018strongly,gonzalez2015deterministic,gonzalez2017efficient,iversen2021strongly,schrinski2022polariton}, the generation of spin squeezing~\cite{Baispin2022}, and the quantum simulation of long-range spin models~\cite{gonzalez2015subwavelength,Douglas_spin_model,Sundaresan,marcoBS,Painter_many_body}.
Relevantly for our purposes, the long-range waveguide-mediated interactions among the emitters lead to 
enhanced collective phenomena as superradiant~\cite{sinha2020non,sierra2022dicke,Cardenas,Arno_super} and subradiant~\cite{AsenjoGarcia_2017,Albrecht_2019,Loic,tiranov2023collective,Laurat,Molmer,dimer_shermet,kumlin2020nonexponential,ostermann2019super,needham2019subradiance,poshakinskiy2021dimerization} emission.
In particular, subradiance effects are characterized by the occurrence of extremely long-lived states that can leak only through the edges of the atomic ensemble.
Motivated by this property and by the  experimental feasibility of waveguide-QED systems, we seek to study the storage of energy in such a platform by answering the following question: ``Can collective effects be used to improve the storage time of a QB?".

To this purpose, we specifically consider a many-body QB embedded within a subset of TLSs coupled to a one-dimensional waveguide, as sketched in Fig.~\ref{fig:sketch}.
We study two different situations. In the first case, the TLSs are equally spaced, forming an ordered array. Here, collective effects are strongly enhanced if the spacing between the emitters is a multiple of half of the atomic resonance wavelength, effectively causing the array to act as a Bragg reflector~\cite{chang2012cavity,corzo2016large,sorensen2016coherent}. Moreover, this system has been considered in a setting similar to ours to study its properties for the transport of energy excitations \cite{Fasser2024}.
However, this scenario has the drawback of being extremely sensitive to disorder in the system.
In the second case, we constructively exploit disorder in the atom spacing by placing the atoms randomly. This setting has been considered in Ref.~\cite{Fayard_2021}, where it was conjectured that disorder may provide many-body localization features to the lattice. We find that these properties are also useful for energy preservation. Specifically, we provide numerical evidence that the long-time behavior of the dissipation follows a power-law decay rather than an exponential one. 

For both configurations, we find that a QB embedded in an emitter ensemble performs better than a one made of a single emitter.
To prove this claim, we performed extensive numerical simulations for the time evolution of the energy, finding that the many-body collective effects help to localize the energy inside the battery, thus protecting it from the environmental dissipation. In addition, in the disordered case, the long-time advantage appears for any lattice spacing, suggesting a structural property of the many-body localized phase. We also evaluate the behavior of the ergotropy~\cite{Hatsopoulos_2, def_ergo, Niedenzu2019, obejko2021}, which accounts for the \textit{useful energy} a battery can store, since it quantifies the maximal extractable work in a closed quantum thermodynamic cycle. Our results show that its evolution is similar to the average energy of the system, certifying the usefulness of the system as a QB.

\section{Model and methods}

In this Section we describe the physical model of our QB. We start with the general Hamiltonian of several TLSs in contact with thermal environment, then we find the correct approximation in the regime we study.
We begin by considering a collection of $L$ TLSs interacting with photons in a one-dimensional semi-infinite waveguide~\cite{Tufarelli_mirror,Baranger_mirror}. The general Hamiltonian is
\begin{eqnarray}
  \nonumber
  \ham & = & \ham_0^L + \sum_{\nu}\int \omega_k\constr_{\nu,k}\destr_{\nu,k} \, {\rm d}k \\
  && + g\sum_{j,\nu}\int \left(\destr_{\nu,k}\exch^je^{i\nu k z_j} + {\rm h.c.} \right) \, {\rm d}k,
  \label{eq:genham}
\end{eqnarray}
where
\begin{equation}
  H_0^L = \omega_0 \sum_{j=1}^L ( I - \sigma_z^j)
  \label{eq:H0}
\end{equation}
is the sum of the local Hamiltonians acting on the TLSs, or atoms [$I$ is the $2\times 2$ identity operator and $\sigma_\alpha^j$ ($\alpha=x,y,z$) the Pauli matrix acting on the $j$th site], with $\omega_0$ the energy of the excited state of the atoms. The $z_j$ variables denote the atom positions, $g$ is the coupling constant between the electromagnetic field and the artificial atoms of the optical lattice, which depends on the dipole moment of the photonic transition, $\exch^j$ ($\dexch^j$) is the raising (lowering) operator of the $j$th site, finally $b_{\nu,k}^{(\dagger)}$ is the annihilation (creation) operator acting on the electromagnetic field with wavevector $k$ at frequency $\omega_k$, while $\nu ={\pm}$ is the propagation direction of the electromagnetic mode.
In this general Hamiltonian, we are considering a continuum of modes. 

In the regime where the system dynamics, ruled by the atomic decay rate, is slower with respect to the time that a photon takes to propagate through the system, we can neglect time delay effects~\cite{carmele2013single,grimsmo2015time,laakso2014scattering,calajo2019exciting,trivedi2021optimal,barkemeyer2021strongly,ask2022non}
and perform a Born-Markov approximation. In this way, we formally integrate out the photonic degrees of freedom, thus obtaining photon-mediated interactions between atoms. Expressed in the interaction picture with respect to the local energy terms of the TLS, the system can be described through the following effective Lindblad master equation (in units of $\hbar=1$)~\cite{AsenjoGarcia_2017, Lalumiere_2013, Albrecht_2019}:
\begin{equation} \label{eq:master}
\frac{d}{dt}\dstate = -i \big( \ham\dstate - \dstate \ham^{\dagger} \big) + \sum_{j,j'}\Gamma_{jj'}\dexch^j\dstate\exch^{j'}
\end{equation}
where $\dstate$ is the density matrix of the TLSs.

A semi-infinite waveguide can be modeled by placing a perfect mirror at $z=0$ so that the non-hermitian Hamiltonian $H$ reads:
\begin{equation}\label{eq:Hsys}
\ham = - i \frac{\Gamma_{1D}}{2} \sum_{j,j'} \big( e^{-ik_{1D}|z_j-z_{j'}|} - e^{-ik_{1D}|z_j+z_{j'}|} \big) \exch^j \dexch^{j'}\,,
\end{equation}
and the collective decay rates are given by \begin{equation}
\Gamma_{jj'} = \Gamma_{1D}[\cos(k_{1D}|z_j-z_{j'}|) - \cos(k_{1D}|z_j+z_{j'}|)].
\end{equation}
Here, $\Gamma_{1D}$ is the spontaneous emission rate into the waveguide, while $k_{1D} = \omega_0/c$ is the wavevector resonant where $c$ is the speed of light in the waveguide.

\begin{figure}[!t]
\centering
\begin{overpic}[width=0.8 \columnwidth]{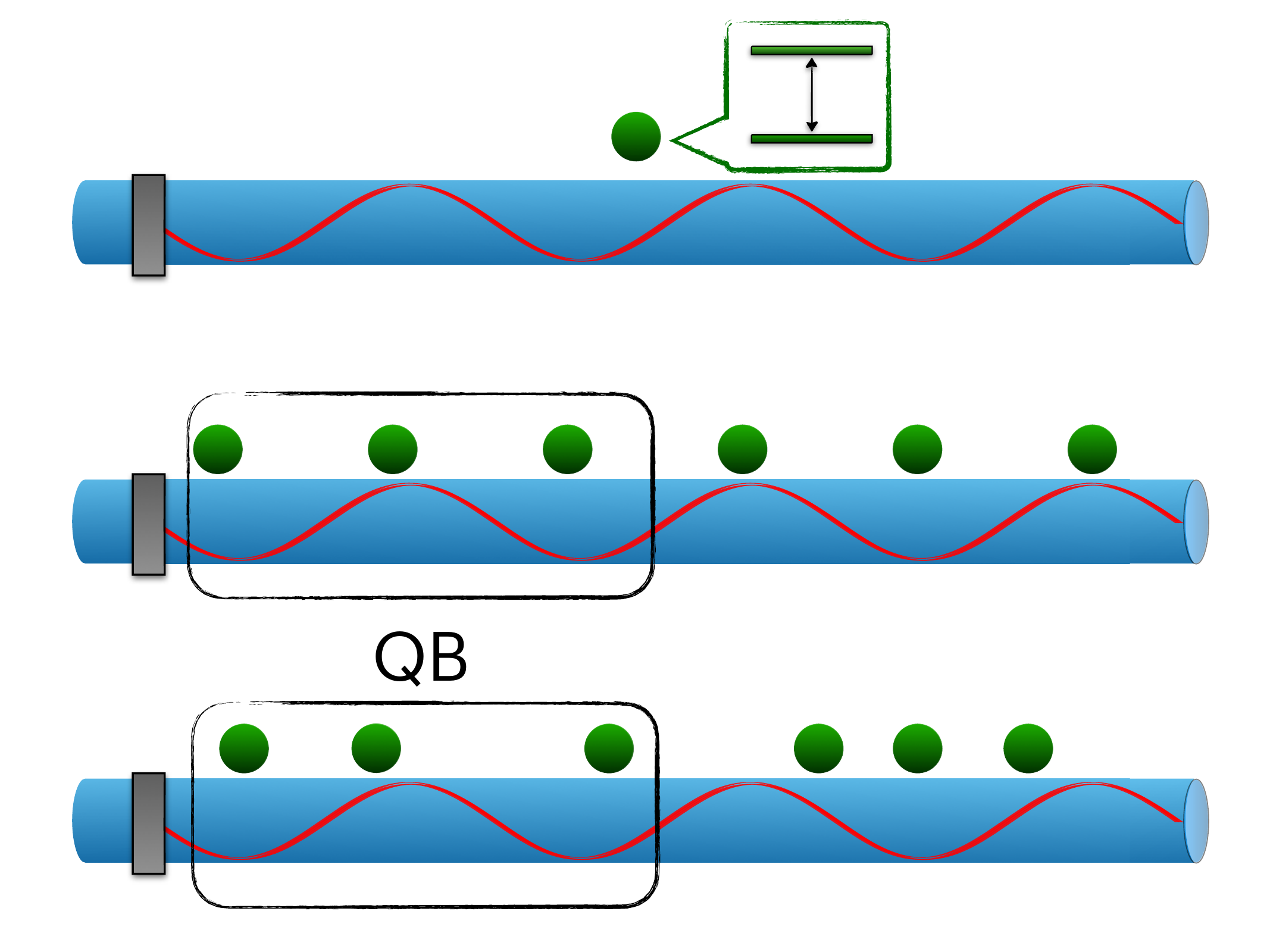}\put(6,68){\normalsize (a) } \put(6,44){\normalsize (b)} \put(6,20){\normalsize (c)}\end{overpic}
\caption{Sketch of our QB setup in three different configurations. Green circles represent two-level atoms coupled to photons of a one-dimensional waveguide, with a reflecting mirror on the left side.
(a) Single-cell QB, with one TLS within the cavity. 
(b) Multi-cell ordered QB, with equally spaced atoms.
(c) Multi-cell disordered QB, with atoms placed at random positions. In (b) and (c) the QB is indicated  by a box, while the remaining TLSs are initially discharged.
\label{fig:sketch}}
\end{figure}

In the following, we focus on three different arrangements of the atoms inside the waveguide (Fig.~\ref{fig:sketch}):
(a) the single-atom ($L=1$) configuration, where collective effects are absent by definition, is adopted as a benchmark for our analysis. 
Then, considering systems with more than one atom ($L>1$), we study:
(b) the ordered scenario, in which the TLSs are regularly spaced inside the cavity at positions $z_j = j d$, where $d$ is the interatomic distance; 
(c) the disordered scenario, where atoms are arranged at positions $z_j=(j+\epsilon_j) d$, where $\epsilon_j \in [-1/2, 1/2]$ is a site-dependent real random variable. 
While for case (a) we can exploit a simple analytic solution, for cases (b) and (c) we resort to a direct numerical integration of the master equation~\eqref{eq:master}, enabling us to access the full density matrix for arbitrarily long times and for systems up to $L=14$ atoms.
In particular, we find it convenient to employ a fourth-order Runge-Kutta integration of the time evolution operator in a vectorized form.
Note that the use of this numerical approach is essential for addressing the subradiant regime, which we aim to utilize for the long-term storage of the QB. This stands in striking contrast to the superradiant regime, where effective mean-field approaches can be applied.\cite{sierra2022dicke,Cardenas}

To protect our system from the environmental noise, we envision a scheme of a QB composed of a number $M<L$ of active sites , i.e., charged TLSs, while the other $L-M$ sites are initially discharged and protect the energy leakage from the cavity. Formally we define the reduced density matrix of the first $M$ cells $\dstate_M(\tau) = \Tr_{L-M}[\dstate(\tau)]$, then we focus on the average local energy on the first $M$ sites:
\begin{equation}
   E_{M,L}(\tau) = {\rm Tr}\big[ {\dstate_M(\tau) \, H_0^M} \big] ,
\end{equation} 
where $H_0^x$ is the free local Hamiltonian describing $x$ identical cells of the QB,
of the same form as in Eq.~\eqref{eq:H0}, and $\tau$ is the duration time of the protocol.
To quantify the fraction of such energy that can be effectively extracted from $M$ cells, without accessing the whole system, we use the ergotropy of the local state $\dstate_M(\tau)$, defined as~\cite{Hatsopoulos_2, def_ergo} 
\begin{equation}
\ergo_{M,L}(\tau) \! = \! \Tr \big[ \dstate_M(\tau)\ham_0^M \big] - \min_{U} \Tr \big[ U \dstate_M(\tau) U^\dagger\ham_0^M \big] ,
\end{equation}
where $U$ is a generic unitary transformation acting on the reduced density matrix. 
Note that, in the disordered scenario, any quantity of interest ${\cal O}$ (e.g., the energy) needs to be averaged over the distribution of the random variables $\{ \epsilon_j \}$. We denote by
\begin{equation}
  \langle \! \langle {\cal O} \rangle \! \rangle= \int P({\epsilon_j}) \, {\cal O} ({\epsilon_j}) \, {\rm d}{\epsilon_j}
\end{equation}
the average quantity, where $P({\epsilon_j})$ is the uniform probability distribution to have a given value of $\epsilon_j \in [-1/2, 1/2 ]$. In the caption of Fig.~\ref{fig:En}, we detail the number of realizations $N_{\rm avg}$ and the parameters of our numerical simulations.

\section{Results}

\begin{figure*}[!t]
\centering
\includegraphics[width=\linewidth]{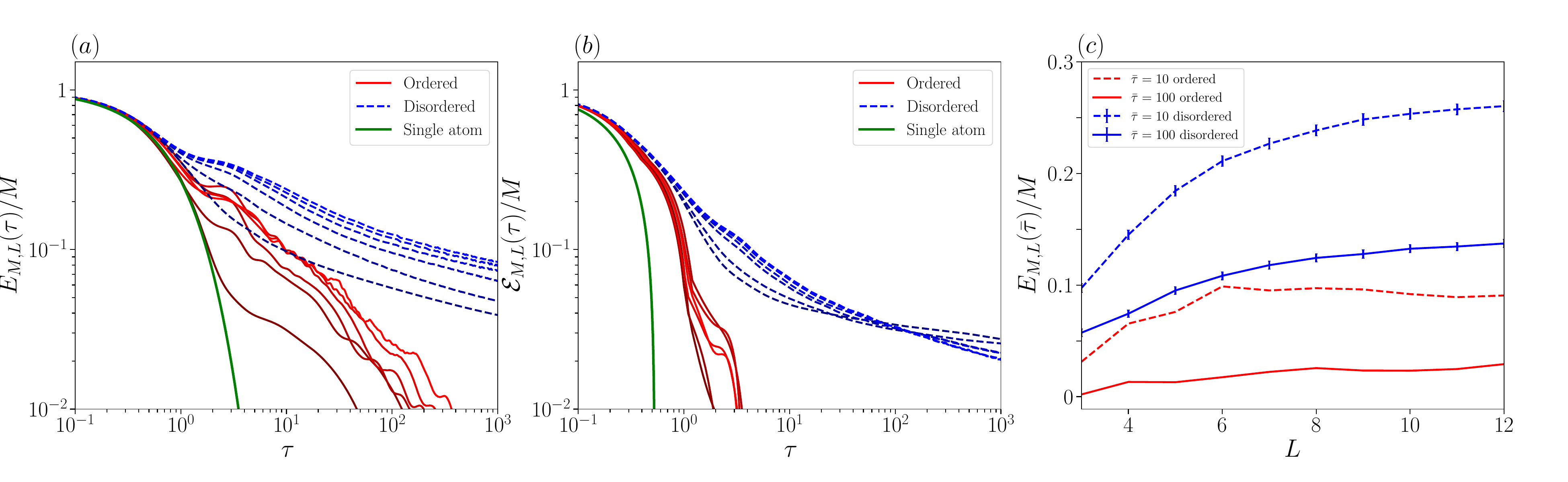}
\caption{(a) The energy stored per quantum cell as a function of the duration time $\tau$, for the ordered case [$E_{M,L}(\tau)/M$, continuous red curves] and for the disordered case after averaging over $10^4$ realizations [$\langle \!\langle E_{M,L}(\tau) \rangle \!\rangle/M$, dashed blue curves], all the energies are rescaled for the input energy of the whole QB, namely $E_{M,L}(0)/M=1$. The various lines are for different numbers of cells, from $L=3$ to $L=9$ and brighten with increasing $L$. Here we always consider an initial state with $M=3$ perfectly charged atoms nearest to the mirror and fix $k_{1D} d = 2.7 \pi$.
(b) Same as in (a), but for the ergotropy $\ergo_{M,L}(\tau)$, rescaled for the initial energy of the battery. The bold green line in each panel is for the corresponding single-atom setup. 
(c) The energy stored as a function of $L$, at fixed time $\bar{\tau}=10$ (dashed curves) and $\bar{\tau}=100$ (continuous curves), for the ordered and the disordered case (red and blue curves, respectively), as taken from (a) and (b); here we also show error bars extrapolated by sampling on the realizations of the disordered model. In all the figures, times $\tau$ are expressed in units of $1/\Gamma_{1D}$.}
\label{fig:En}
\end{figure*}

We initially charge the $M$ cells nearest to the mirror. This means that, at time $\tau=0$, such $M$ TLSs are perfectly excited, while all the others are in their ground state.
Then we let the whole system naturally evolve through the dynamics provided by the master equation~\eqref{eq:master}: this represents the storage protocol we are focusing on. Most importantly, we are going to show that a many-body QB model performs better than a single TLS loaded in a semi-infinite optical waveguide. 
In particular, it is known that the collective behavior of multiple emitters coupled to a waveguide can lead to a subradiant decay of an initial excited state~\cite{Albrecht_2019,Loic,dimer_shermet}. 
Here we consider the thermodynamical consequences of this feature, which allow us to build a resilient QB.

First, we consider the performance of a single atom in a cavity as our benchmark. In this scenario ($L=1$), the average local energy and the ergotropy are expected to decay exponentially fast in time. Starting from an initially excited configuration $\rho(0) = \ket{1}\!\!\bra{1}$, the TLS will decay into the one-dimensional electromagnetic environment at a rate that depends on the wavevector $k$, i.e.,  $\Gamma_{0} = \Gamma_{1D}[1-\cos(k_{1D}|2d|)]$. In fact, the time evolution of its reduced density operator reads
\begin{equation}
\rho(\tau) = \begin{pmatrix}
  1-e^{-\Gamma_{0}\tau} & 0 \\
  0 & e^{-\Gamma_{0}\tau}
\end{pmatrix}.
\end{equation}
From this expression, it is straightforward to compute the average energy of the atomic system in time, given by
\begin{equation}
  E_{\rm 1-TLS}(\tau) = \Tr \big[ \rho(\tau) \, H_0^1 \big] = \omega_0e^{-\Gamma_0\tau},
\end{equation}
while the corresponding ergotropy reads
\begin{equation}
    \ergo_{\rm 1-TLS}(\tau) = \left\{\begin{array}{cc}
       \omega_0(2e^{-\Gamma_{0} \tau} - 1) & \;\; \mbox{for } \; \Gamma_{0} \tau < \ln 2 , \vspace*{1mm}\\
       0  & \;\; \mbox{for } \; \Gamma_{0} \tau \geq \ln 2 .
    \end{array}\right.
\end{equation}
In contrast, as detailed in the following, for $L>1$ we find a substantial quantitative improvement, due to a collective slower relaxation.

\begin{figure}[h!]
    \centering
    \includegraphics[width=1.1\linewidth]{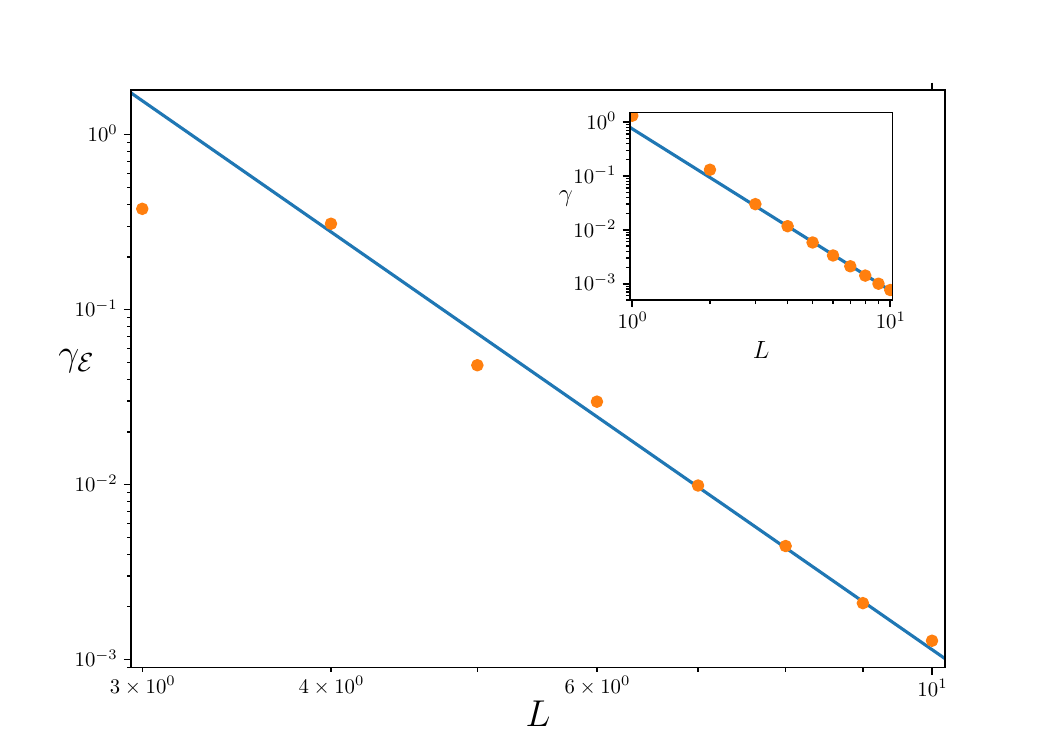}
    \caption{In the main plot we show the temporal decay rate $\gamma_{\ergo}$ of the ergotropy. Orange circles are obtained by exponentially fitting the $\ergo_{M,L}(\tau)$ data at fixed $L$, while the straight blue line is a power-law fit $\gamma_\ergo = B/L^6$, with $B \approx 1139.2$. In the inset is plotted the temporal decay rate $\gamma$ of the energy at long times, in the ordered scenario. Orange circles denote the numerical results of an exponential fit for the $E_{M,L}(\tau)$ data at fixed $L$, while the straight blue line represents the power-law fitting function $\gamma = A/L^3$, with $A \approx 0.747$. In both plots we use a log-log scale, set $k_{1D}d = 2.7 \pi$, $M=3$, and the quantities $E_{M,L}(\tau)$ and $\ergo_{M,L}(\tau)$ are rescaled for the input energy of the QB. In all plots  both $\gamma$ and $\gamma_{\ergo}$ are computed in units of $\Gamma_{1D}$.}
    \label{fig:rate}
\end{figure}

\begin{figure}[!t]
\centering
\includegraphics[width=\linewidth]{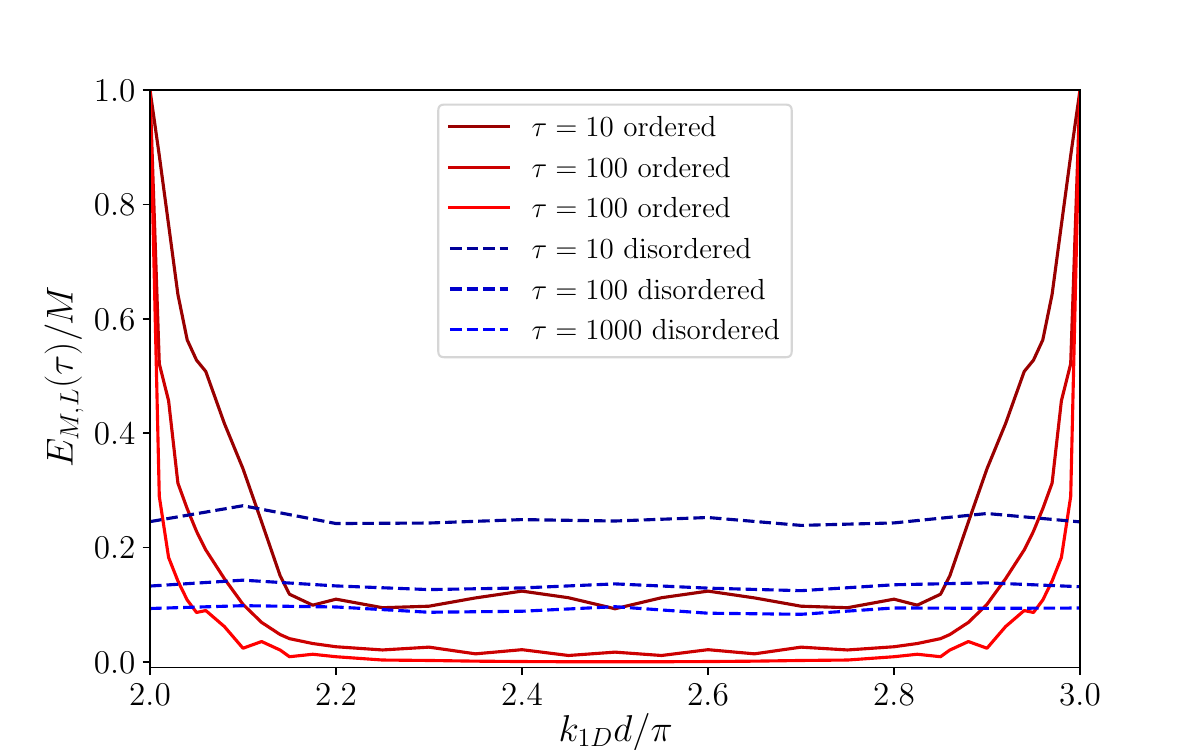}
\caption{The energy stored as a function of the average distance $k_{1D} d / \pi$ between the atoms, for the ordered case ($E_{M,L}/M$, continuous red curves) and for the disordered case ($\langle\!\langle E_{M,L} \rangle \!\rangle/M$, dashed blue curves). Here all the energies are rescaled for the input energy of the QB. Each of the various lines is for a fixed time $\tau = 10, \, 10^2, \, 10^3$ (in units of $\Gamma_{1D}^{-1}$), with brightening color when increasing $\tau$.
Here we fix $L=8$, $M=3$.
\label{fig:EVsk}}
\end{figure}

Figure~\ref{fig:En} displays (a) the average energy $E_{M,L}(\tau)$ and (b) the ergotropy $\ergo_{M,L}(\tau)$, normalized per atom, of the $M=3$ initially charged atoms, as functions of the storage time $\tau$. Both the ordered and the disordered system (red vs.~blue data sets) undergo a transient phase in which part of the internal energy is lost. In panel (c) we show the energy remaining in the battery after two different times $\tau$, for both the ordered and the disordered case.
The initial energy loss is comparable to the decay of an individual atom (benchmark bold green curve). After some time, the ordered system shows an exponential decay, with a rate $\sim \Gamma^{(1)}_1$ mainly given by the most subradiant collective single excitation state~\cite{Loic}, while the behavior of the disordered system appears to be power-law at long times. 
This decay in the ordered case is known to depend on the inter-atomic distance $k_{1D}d/\pi$, as shown in Fig.~\ref{fig:EVsk}, and on the size of the system at fixed initial configuration, decaying as $\gamma \propto L^{-3}$~\cite{Albrecht_2019}.  
Note that for larger battery sizes, $M\gtrsim 20$, the system should start to exhibit a power-law decay in time, making this configuration more stable~\cite{Loic}, as shown in the inset of Fig.~\ref{fig:rate}. Let us remark that the chosen initial state does not favor either superradiance nor subradiance; after an intial superradiant  transient phase, we observe that the subradiant process becomes predominant.
In addition, the ergotropy in the ordered scenario also presents an exponential temporal behavior at very long times, although with a decay rate that curiously appears to have a different power-law dependence on the system size, as $\gamma_{\ergo} \propto L^{-6}$. This behavior is displayed in the main plot of Fig.~\ref{fig:rate}.

On the other hand, after the transient behavior, the disordered system displays a much slower decay, similar to a power law. This qualitatively different functional decay at long times allows us to preserve the remaining energy for much longer times. Moreover, the energy of the charged sites is progressively more protected as the system size $L$ is increased. 
Something qualitatively analogous occurs to the ergotropy [see Fig.~\ref{fig:En}(b)], although in the ordered scenario we observe dips and revivals that are linked to the fact that, when the state is losing energy, after a certain time it becomes purer so that a larger fraction of its energy can be extracted. We do not show such dips, since they occur at values of the ergotropy smaller than $10^{-2}$, and thus are of no practical value.

\begin{figure}[!t]
    \centering
    \includegraphics[width=\linewidth]{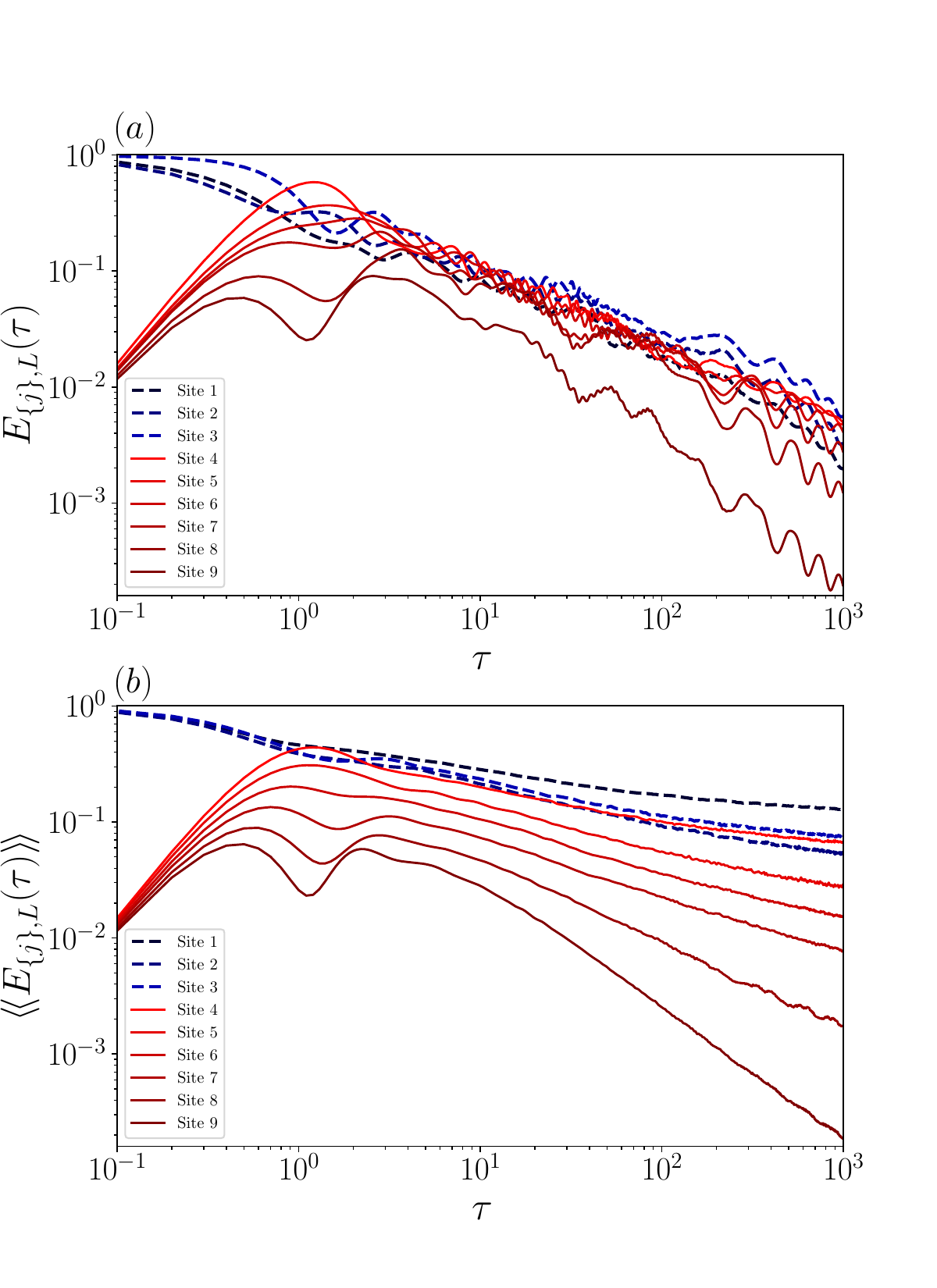}
    \caption{(a) The local energy $E_{\{j\},L}(\tau)$, for each site $j$ of the chain, as a function of the evolution time $\tau$, in the ordered case. (b) The mean local energy $\langle \!\langle E_{\{j\},L}(\tau) \rangle \!\rangle$, after averaging over $10^4$ realizations, for each site $j$ of the chain, as a function of $\tau$, in the disordered case.
    We consider a system of size $L=9$ and load an initial state with $M=3$ perfectly charged atoms nearest to the mirror. The local input energy of each charged site is $E_{\{j\},L}(0)=1$ for $j=1,2,3$, and we fix $k_{1D} d = 2.7 \pi$.}
    \label{fig:en_loc}
\end{figure}

To better understand the mechanism lying at the basis of energy preservation with disorder, we have also considered the local energy for each TLS. Namely, Fig.~\ref{fig:en_loc} shows the time behavior of
\begin{equation}
  E_{\{j\},L}(\tau) = \Tr \big[ \rho_j(\tau) \, (I-\sigma_z^j) \big], \quad (j=1,\ldots,L),
\end{equation}
where $\rho_j(\tau) = {\rm Tr}_{L-\{j\}} [\rho(\tau)]$ is the reduced density matrix
corresponding to the $j$th site.
We considered a system with $L=9$ and the first $M=3$ sites initially charged.
We observe that, in the ordered case (a), the discharged sites pick up energy over time and the system eventually loses energy exponentially in time. In contrast, in the disordered case (b), the first three sites retain a larger fraction of the input energy compared to the ordered case, while the other sites do not get fully charged in time. Thus, a consistent fraction of the energy remains localized in the charged sites freezing the energy dynamics in time. The size of the energy spreading depends on the localization length of the disordered model. This energy-preserving behavior can be compared to the findings of Ref.~\cite{Fayard_2021}, where the authors show that the disordered arrangement of the atoms inside the cavity shows evidence of a many-body localized phase.

Finally in Fig.~\ref{fig:EVsk} we study the quality of the system as a QB by tuning the parameter $k_{1D}d/\pi$: if the atoms are arranged in an ordered way, there are special points in which the decay rate goes to zero, i.e., when $k_{1D}=n\pi/d$ 
for any $n \in \mathbb{N}$. On the other hand, in the presence of spatial disorder, this parameter becomes irrelevant. This proves that collective phenomena enhance the energy-storing properties of QBs, because both ordered and disordered cases outperform the single-atom benchmark at long times. However, while in the case the TLSs are arranged periodically the most favorable condition is fragile (since we need a great amount of control on the parameter $k_{1D}d/\pi$), in the disordered case one is able to build a resilient QB in a robust way. 

\section{Discussion}

We have analyzed two different arrangements of a one-dimensional waveguide-QED system: an ordered and a disordered array. We performed extensive numerical simulations of the time evolution for these two systems and compared the energy decay to a single-atom benchmark.
In the ordered case, the protection properties were enhanced by many-body effects of the system, but only at specific lattice spacings where the optical lattice behaves as an insulator. In contrast, in the disordered case, we discovered that, regardless of the parameters chosen, the system protects a fraction of the energy and ergotropy within the lattice for extended periods.
Therefore, we demonstrated how collective waveguide-QED effects can enhance energy storage performances.
Let us stress that in this study we analyzed a waveguide instead of a cavity, because on the one hand it gives a channel to charge the battery without changing the structure of the system. On the other hand, it models the interaction between the battery and the environment more realistically; in fact, in this model, we consider energy leaking from the lattice boundary.
We notice, however, that these many-body effects may be detrimental in the charging and discharging phase, because they induce a strong energetic rigidity in the system, which is essential during the storage, but it might be a hindrance to change the charge of the QB. Thus, a fully operational QB should ideally exhibit two distinct operational phases: one dedicated to the efficient charging and discharging, and another one optimized for the long-term energy storage.

In the future, it would be tempting to experimentally test our predictions using atoms in a waveguide~\cite{roy_rev_mod, Darrick_rev_mod} or with molecular excitons~\cite{tibben2024extendingselfdischargetimedicke}. In realistic implementations, the coupling to modes outside the waveguide would rapidly compromise the QB storage capacity. While this issue is relatively less important in circuit-QED platforms~\cite{Astafiev2010,painter2,Kannan,ustinov}, it becomes critical in the optical regime, where only a small fraction of the atomic emission couples into the waveguide. One potential solution to this challenge involves using sub-wavelength atomic arrays, where collective free-space emission could be engineered to preferentially radiate into the waveguide via the so called ``selective radiance" mechanism~\cite{AsenjoGarcia_2017}.  
A QB is a complex quantum device where energy can be injected, stored, and finally extracted. Therefore, it is important to interface our findings, where collective effects enhance energy storage, with cases where collective effects are used to boost charging \cite{PhysRevApplied.14.024092, rossini2019quantum, Gyhm2022}. In particular, we remark that waveguide-QED systems are capable of being used in any of the phases of a QB. In fact, the charging phase can be carried out through by driving the system to an excited state. The storage phase is detailed in this article. Finally, the QB discharging (i.e., the work extraction) may be performed by changing in time the couplings of the model, or, alternatively, by tuning the frequencies of the prescribed Hamiltonian.

\acknowledgments

We thank M. Polini and D. E. Chang for useful discussions and insightful suggestions.
S.T. acknowledges financial support by MUR (Ministero dell’ Universit\`a e della Ricerca) through the following projects: 
PNRR MUR project PE0000023-NQSTI, PRIN 2017 Taming complexity via Quantum Strategies: a Hybrid Integrated Photonic approach (QUSHIP) Id.~2017SRN-BRK, and project PRO3 Quantum Pathfinder.
G.M.A. acknowledges funding from the European Research Council (ERC) under the European Union's Horizon 2020 research and innovation program (Grant agreement No.~101002955 -- CONQUER).
G.C. acknowledges that results incorporated in this standard have received funding from the T-NiSQ consortium agreement financed by QUANTERA 2021 and that this work is supported in part by the Italian MUR Departments of Excellence grant 2023-2027 ``Quantum Frontiers".


%

\end{document}